\documentclass[thmsa]{article}
\usepackage{sw20lart}

\input{tcilatex}
\begin{document}

\title{n-p Interaction Effects on the Double Beta Decay Nuclear Matrix Elements for
Medium Mass Nuclei}
\date{}
\author{P. K. Raina and A. Shukla \\
\textit{Department of Physics and Meteorology, IIT Kharagpur-721302, India}
\and P. K. Rath, B. M. Dixit, K. Chaturvedi, and R. Chandra. \\
\textit{Department of Physics, University of Lucknow, Lucknow-226007, India}
\and S. K. Dhiman and A. J. Singh. \\
\textit{Department of Physics, H. P. University, Shimla-171002, India.}}
\maketitle

\begin{abstract}
\-The quality of HFB wave functions are tested by comparing the
theoretically calculated results with the available experimental data for a
number of spectroscopic properties like yrast spectra, reduced B(E2)
transition probabilities, quadrupole moments and g-factors for the nuclei
involved in 2$\nu $ $\beta \beta $ decay. It is observed that the np
interactions vis-\`{a}-vis the deformations of the intrinsic ground states
of medium mass nuclei play a crucial role in the fine tuning of the nuclear
matrix elements, M$_{2\nu }.$
\end{abstract}

\-It is well established by now that the implications of nuclear $\beta
\beta $ decays are far reaching in nature in general. 0$\nu $ $\beta \beta $
decay in particular is one of the very rare promising processes to test the
physics beyond the standard model (SM) of fundamental particles. These
aspects of nuclear $\beta \beta $ decay have been excellently elaborated in
a number of review articles over the past years [1-8].

The 2$\nu $ $\beta \beta $ decay, a second order process of weak interaction
that conserves the lepton number exactly, is allowed in the SM. The half
life of 2$\nu $ $\beta \beta $ decay is a product of accurately known phase
space factor and appropriate nuclear transition matrix element M$_{2\nu }$.
The half lives of 2$\nu $ $\beta \beta $ decay have been already measured
for about ten nuclei and the values of M$_{2\nu }$ can be extracted
directly. Consequently, the validity of different models employed for
nuclear structure calculations can be tested by calculating the M$_{2\nu }$.

It is observed that in all cases the 2$\nu $ $\beta \beta $ decay matrix
elements are sufficiently quenched. The main motive of all the theoretical
calculations is to understand the physical mechanism responsible for the
suppression of the M$_{2\nu }$. The M$_{2\nu }$ is calculated mainly in
three types of models. One is the shell model and its variants. The second
is the quasiparticle random phase approximation (QRPA) and extensions their
of. The third type of models is classified as the alternative models. The
details about these models - their advantages as well as shortcomings - have
been discussed excellently by Suhonen and Civitarese [5] and Faessler and
Simkovic [6].

All the nuclei undergoing $\beta \beta $ decay are even-even type. Hence the
pairing degrees of freedom play an important role. Moreover, it has been
conjectured that the deformation can play a crucial role in $\beta \beta $
decay rates. Hence it is desirable to have a model that incorporates the
pairing and deformation degrees of freedom on equal footing in its
formalism. For this purpose the Projected Hartree-Fock-Bogoliubov (PHFB)
model is one of the most natural choices. Coincidentally, most of the $\beta
\beta $ decaying nuclei fall in medium mass region. The success of the PHFB
model in explaining the observed experimental trends in this mass region has
motivated us to apply the PHFB wave functions to the study of nuclear $\beta
\beta $ decay as well.

The mass region A$\approx $100 provides us with a nice example of shape
transitions [9], where, at one end, nuclei can be described in terms of
shell model wave functions involving a small number of configurations and,
at the other end of this region, we find good evidence of rotational
collectivity. These nuclei lie between doubly magic $^{132}$Sn and the
strongly deformed $^{100}$Zr, near which the structural changes are rather
rapid with the addition of protons and neutrons. In the past there have been
many attempts, [10-15] to explore the factors responsible for the structural
changes in this mass region.

\smallskip Federman and Pittel [14] computed the deformation energy in the
framework of Hartree-Fock-Bogoliubov (HFB) theory in conjunction with the
surface delta interaction (SDI), suggesting that the neutron-proton (n-p)
interaction in the spin-orbit partner (SOP) orbits-1g$_{^{9/2}}$ and 1g$%
_{7/2}$ in this case may be instrumental vis-\`{a}-vis the onset of
deformation in Mo isotopes with A\TEXTsymbol{>}100. A systematic study of
the behavior of the low-lying collective states of neutron-rich even Cd, Pd,
Ru, and Mo isotopes has lead to the conclusion that these structural changes
are related to the exceptionally strong n-p interaction in this region. It
has also been observed that the n-p interactions among the SOP orbits have a
deformation producing tendency and the systematics of low-lying states are
intricately linked with the nature of n-p interaction.

\smallskip The sensitivity of the yrast spectra and the transition charge
densities (TCD) to the neutron-proton interaction strength has lead to the
fixing of these strengths very accurately and has been demonstrated [15]
through the examples of $^{110}$Cd and $^{114}$Cd. We have adopted this
method for fixing the p-n strength of QQ interaction by looking at the
spectra of 2$^{+}$ state of the nuclei involved in $\beta \beta $ decay.

A large number of theoretical as well as experimental studies of 2$\nu $ $%
\beta \beta $ decay have already been done for $\beta ^{-}\beta ^{-}$ of $%
^{96}$Zr, $^{100}$Mo, $^{110}$Pd, $^{128,130}$Te nuclei and e$^{+}$ $\beta
\beta $ ( $\beta ^{+}\beta ^{+}$, $\beta ^{+}$EC and ECEC) in case of $^{96}$%
Ru, $^{106}$Cd, $^{124}$Xe and $^{130}$Ba nuclei over the past few years
with more emphasis on $^{100}$Mo and $^{106}$Cd cases. The $\beta \beta $
decay is not an isolated nuclear process. The availability of data permits a
rigorous and detailed critique of the ingredients of the microscopic models
used to provide a description of these nuclei.

We have studied the 2$\nu $ $\beta \beta $ decay not isolatedly but together
with other observed nuclear phenomena. This is in accordance with the basic
philosophy of nuclear many body theory, which is to explain all the observed
properties of nuclei in a coherent manner. Hence as a test of the
reliability of the wave functions, we have calculated the yrast spectra,
reduced B(E2) transition probabilities, static quadrupole moments and
g-factors and compared with the available experimental data.

The theoretical formalism to calculate the half life of 2$\nu $ $\beta \beta 
$ decay mode has been given by Haxton and Stephenson [1] , Doi et al [2,3]
and Tomoda [4]. Very brief outlines of the calculation of nuclear transition
matrix elements of the $\beta \beta $ decay in the PHFB model are presented
here. Details of expressions used in calculation of spectroscopic properties
can be found in Dixit et al [16].

The half-life of 2$\nu $ $\beta \beta $ decay for 0$^{+}\rightarrow $0$^{+}$
transition is given by

\begin{equation}
\left[ T_{1/2}^{2\nu }(0^{+}\longrightarrow 0^{+})\right] ^{-1}=G_{2\nu
}\left| M_{2\nu }\right| ^{2}
\end{equation}
where\smallskip 
\begin{equation}
M_{2\nu }=\sum_{N}\frac{\left\langle 0^{+}\left\| \mathbf{\sigma }\tau
^{+}\right\| 1_{N}^{+}\right\rangle \left\langle 1_{N}^{+}\left\| \mathbf{%
\sigma }\tau ^{+}\right\| 0^{+}\right\rangle }{E_{N}-(M_{I}+M_{F})/2}
\end{equation}
and the integrated kinematical factor G$_{2\nu }$ can be calculated with
good accuracy [8]. If the E$_{N}$ of Eq.(11) is replaced by an average $%
\left\langle E_{N}\right\rangle $, the summation over intermediate states
can be completed using the closure approximation one obtains

\begin{equation}
M_{2\nu }=-\frac{2M_{GT}^{2\nu }}{\left\langle E_{N}\right\rangle -\left(
M_{I}+M_{F}\right) /2}=-\frac{2M_{GT}^{2\nu }}{E_{d}}
\end{equation}
where the double Gamow-Teller matrix element (DGT) $M_{GT}^{2\nu }$ is
defined as follows

\begin{equation}
M_{GT}^{2\nu }=\frac{1}{2}\left\langle 0^{+}\left| \sum_{n,m}\mathbf{\sigma }%
_{n}.\mathbf{\sigma }_{m}\tau _{n}^{+}\tau _{m}^{+}\right| 0^{+}\right\rangle
\end{equation}

Employing the HFB wave functions, one obtains the following expression for
the $\beta \beta $ decay nuclear transition matrix element.

\begin{eqnarray}
\left\langle M_{GT}^{2\nu }\right\rangle 
&=&[n_{Z-2,N+2}^{J_{f}=0}n_{Z,N}^{J_{i}=0}]^{-1/2}\int\limits_{0}^{\pi
}n_{(Z,N),(Z-2,N+2)}(\theta )  \nonumber \\
&&\times \frac{1}{4}\sum_{\alpha \beta \gamma \delta }\left\langle \alpha
\beta \left| \mathbf{\sigma }_{1}.\mathbf{\sigma }_{2}\tau ^{+}\tau
^{+}\right| \gamma \delta \right\rangle \sum_{\varepsilon \eta }\left[
\left( 1+F_{Z,N}^{(\nu )}(\theta )f_{Z-2,N+2}^{(\nu )}\right) \right]
_{\varepsilon \alpha }^{-1}(f_{Z-2,N+2}^{(\nu )})_{\varepsilon \beta } 
\nonumber \\
&&\times \left[ \left( 1+F_{Z,N}^{(\pi )}(\theta )f_{Z-2,N+2}^{(\pi
)}\right) \right] _{\gamma \eta }^{-1}(F_{Z,N}^{(\pi )})_{\eta \delta }\sin
\theta d\theta   \label{eq1}
\end{eqnarray}
where

\begin{equation}
n^{J}=\int\limits_{0}^{\pi }\{\det [1+F^{(\pi )}(\theta )f^{(\pi )\dagger
}]\}^{1/2}\times \{\det [1+F^{(\nu )}(\theta )f^{(\nu )\dagger
}]\}^{1/2}d_{00}^{J}(\theta )\sin (\theta )d\theta 
\end{equation}
and

\begin{equation}
n_{(Z,N),(Z-2,N+2)}(\theta )=\{\det [1+F_{Z,N}^{(\pi )}(\theta
)f_{Z-2,N+2}^{(\pi )\dagger }]\}^{1/2}\times \{\det [1+F_{Z,N}^{(\nu
)}(\theta )f_{Z-2,N+2}^{(\nu )\dagger }]\}^{1/2}
\end{equation}

\smallskip

The $\pi $ ($\nu $) represents the proton (neutron). PHFB calculations are
summarized by the coefficients (U$_{im}$,V$_{im}$) and C$_{ij,m}$ and their
matrices (F$_{N,Z}$($\theta $))$_{\alpha \beta }$and (f$_{N,Z}$). The
details can be found in reference [16].

In the present calculations we treat the doubly even nucleus $^{76}$Sr
(Z=N=38) as an inert core and the valence space is spanned by the orbits 1p$%
_{1/2}$, 2s$_{1/2}$, 1d$_{3/2}$, 1d$_{5/2}$, 0g$_{7/2}$, 0g$_{9/2}$ and 0h$%
_{11/2}$ for protons and neutrons. The set of single particle energies
(SPE's) but for the $\varepsilon $(0h$_{11/2}$) which is slightly lowered,
employed here is same as used in a number of successful shell model as well
as variational model [10-16] calculations for nuclear properties in the mass
region A=100. The effective two-body interaction is the PPQQ type [17].

\textit{Table 1. Variation in excitation energies in MeV of J}$^{\pi }$%
\textit{=2}$^{+}$\textit{, 4}$^{+}$\textit{\ and 6}$^{+}$\textit{\ yrast
states for }$^{100}$Mo\textit{\ and }$^{100}$Ru\textit{\ nuclei with change
in }$\chi _{pn}$\textit{\ keeping fixed G}$_{p}$\textit{\ = -0.30 MeV, G}$%
_{n}$\textit{\ = -0.20 MeV and }$\varepsilon $\textit{(0h}$_{11/2}$\textit{)
= 8.6 MeV.\smallskip }

\begin{tabular}{llllllll}
\hline\hline
Nucleus & $\chi _{pn}$ & 0.01826 & 0.01866 & 0.01906 & 0.01946 & 0.01986 & 
Exp.[21] \\ \hline
& E$_{2+}$ & 0.6865 & 0.5851 & 0.5356 & 0.4493 & 0.3923 & 0.5355 \\ 
$^{100}$Mo & E$_{4+}$ & 1.7028 & 1.5333 & 1.4719 & 1.3070 & 1.1861 & 1.1359
\\ 
& E$_{6+}$ & 2.9355 & 2.7213 & 2.6738 & 2.4560 & 2.2854 &  \\ \hline
& $\chi _{pn}$ & 0.01758 & 0.01798 & 0.01838 & 0.01878 & 0.01918 &  \\ \hline
& E$_{2+}$ & 0.6597 & 0.5923 & 0.5395 & 0.4930 & 0.4445 & 0.5396 \\ 
$^{100}$Ru & E$_{4+}$ & 1.8175 & 1.6733 & 1.5591 & 1.4531 & 1.3372 & 1.2265
\\ 
& E$_{6+}$ & 3.2746 & 3.0615 & 2.8940 & 2.7329 & 2.5519 & 2.0777 \\ 
\hline\hline
\end{tabular}

\smallskip

The strength of the pairing interaction is fixed through the relation G$_{p}$
= -30/A MeV and G$_{n}$= -20/A MeV. These values of G$_{p}$ and G$_{n}$ have
been used by Heestand et al [18] to successfully explain the experimental g(2%
$^{+}$) data of some even-even Ge, Se, Mo, Ru, Pd, Cd and Te isotopes in
Greiner's collective model [19]. The strengths of the like particle
components of the QQ interaction are taken as: $\chi _{pp}$ = $\chi _{nn}$ =
-0.0105 MeV b$^{-4}$. These values for the strength of the interaction are
comparable to those suggested by Arima on the basis of an empirical analysis
of the effective two-body interactions [20].

Table 2.\textit{\ Experimental half-lives T}$_{1/2}^{2\nu }$\textit{, and
corresponding nuclear matrix elements M}$_{2\nu }$\textit{, along with the
theoretical values in different models for }$0^{+}\rightarrow 0^{+}$\textit{%
\ 2}$\nu \beta \beta $\textit{\ decay of }$^{100}$Mo\textit{. The numbers
corresponding to (a) and (b) are calculated for g}$_{A}$\textit{=1.25 and
1.0 respectively.}

\begin{tabular}{llcc|lllll}
\hline\hline
\multicolumn{4}{c|}{Experiment} & \multicolumn{5}{c}{Theory} \\ \hline
{\small Ref} & {\small Projects} & {\small T}$_{1/2}^{2\nu }$ & $\left|
M_{2\nu }\right| $ & {\small Ref} & {\small Models} & $\left| M_{2\nu
}\right| $ & \multicolumn{2}{c}{{\small T}$_{1/2}^{2\nu }$} \\ 
&  & {\small \ }$(10^{18}yrs)$ & \multicolumn{1}{l|}{} &  &  &  & 
\multicolumn{2}{c}{$(10^{18}yrs)$} \\ 
&  & \multicolumn{1}{l}{} & \multicolumn{1}{l|}{} &  &  &  & {\small a)} & 
{\small b)} \\ \hline
{\small \lbrack 22]} & {\small UC-Irvin} & \multicolumn{1}{l}{{\small 6.82}$%
_{-0.53}^{+0.38}\pm 0.68$} & \multicolumn{1}{l|}{{\small a)0.125}$%
_{-0.009}^{+0.012}$} & {\small [16]} & \multicolumn{1}{c}{\small PHFB} & 
{\small 0.152} & {\small 4.57} & {\small 11.15} \\ 
&  & \multicolumn{1}{l}{} & \multicolumn{1}{l|}{{\small b)0.195}$%
_{-0.020}^{+0.014}$} & {\small [24]} & \multicolumn{1}{c}{\small SRPA(WS)} & 
{\small 0.059} & {\small 30.4} & {\small 74.3} \\ 
{\small \lbrack 23]} & {\small NEMO} & \multicolumn{1}{l}{{\small 9.5}$\pm
0.4\pm 0.9$} & \multicolumn{1}{l|}{{\small a)0.106}$_{-0.007}^{+0.008}$} & 
{\small [25]} & \multicolumn{1}{c}{\small SU3(SPH)} & {\small 0.152} & 
{\small 4.59} & {\small 11.2} \\ 
&  & \multicolumn{1}{l}{} & \multicolumn{1}{l|}{{\small b)0.165}$%
_{-0.010}^{+0.013}$} & {\small [25]} & \multicolumn{1}{c}{\small SU3(DEF)} & 
{\small 0.108} & {\small 9.09} & {\small 22.2} \\ 
{\small \lbrack 8]} & {\small Average} & \multicolumn{1}{l}{{\small 8.0}$\pm
0.7$} & \multicolumn{1}{l|}{} & {\small [26]} & \multicolumn{1}{c}{\small %
QRPA(EMP)} & {\small 0.197} & {\small 2.73} & {\small 6.67} \\ \hline\hline
\end{tabular}

\smallskip

As an illustrative case we look into the details of calculations for double
beta decay of $^{100}$Mo nucleus. The $\chi _{pn}$ is varied so as to obtain
the spectra of $^{100}$Mo and $^{100}$Ru in optimum agreement with the
experimental results. In Table 1, we have presented the theoretically
calculated yrast energies for levels of $^{100}$Mo and $^{100}$Ru for
different values of $\chi _{pn}$. It is clearly observed that as the $\chi
_{pn}$ is varied by 0.0016 MeV b$^{-4}$, the E$_{2},$ decreases by 0.2942
MeV in case of $^{100}$Mo and 0.2152 MeV in case of $^{100}$Ru respectively.
This is understandable as there is an enhancement in the collectivity of the
intrinsic state with the increase of \TEXTsymbol{\vert}$\chi _{pn}|$, hence
the E$_{2}$ decreases. The optimum values of $\chi _{pn}$ corresponding to $%
^{100}$Mo and $^{100}$Ru are 0.01906 MeV b$^{-4}$ and -0.01838 MeV b$^{-4}$
respectively. Thus for a given model space, SPE's, G$_{p}$, G$_{n}$ and $%
\chi _{pp}$, we have fixed $\chi _{pn}$ through the experimentally available
energy spectra.

\textit{Table 3. Experimental limit on half-lives (T}$_{1/2}^{2\nu }$\textit{%
) and corresponding extracted matrix elements M}$_{2\nu }$\textit{\ along
with their theoretically calculated values for 2}$\nu $\textit{\ }$\beta
^{+}\beta ^{+}$\textit{\TEXTsymbol{\backslash}}$\beta ^{+}$\textit{EC%
\TEXTsymbol{\backslash}ECEC decay of }$^{106}$Cd\textit{\ for 0}$%
^{+}\rightarrow $\textit{0}$^{+}$\textit{\ transition. The numbers
corresponding to (a) and (b) are calculated for g}$_{A}$\textit{=1.25 and
1.0 respectively.}

\smallskip 
\begin{tabular}{l|ll|lllll}
\hline\hline
Decay & \multicolumn{2}{|c|}{$Experiment$} & \multicolumn{5}{|c}{Theory} \\ 
\cline{2-8}
Mode$^{\dagger }$ & Ref & T$_{1/2}^{2\nu }(yrs)$ & Ref & Model & $\left|
M_{2\nu }\right| $ & \multicolumn{2}{c}{(T$_{1/2}^{2\nu })^{\dagger }$} \\ 
&  &  &  &  &  & a) & b) \\ \hline
$\beta ^{+}\beta ^{+}$ & [27] & \TEXTsymbol{>}2.4$\times 10^{20**}$ & [31] & 
PHFB & 0.238 & 35.42 & 89.56 \\ 
10$^{25}$ yrs & [28] & \TEXTsymbol{>}1.0$\times 10^{19*}$ & [32] & QRPA & 
0.166 & 72.79$^{\#}$ & 180$^{\#}$ \\ 
& [29] & \TEXTsymbol{>}9.2$\times 10^{17}$ & [28] & QRPA & 1.226 & 1.33 & 3.3
\\ \cline{1-3}\cline{2-8}
$\beta ^{+}EC$ & [27] & \TEXTsymbol{>}4.1$\times 10^{20}$ & [31] & PHFB & 
0.238 & 8.97 & 22.69 \\ 
10$^{21}$yrs & [28] & \TEXTsymbol{>}0.66$\times 10^{19*}$ & [32] & QRPA & 
0.169 & 17.79$^{\#}$ & 44.0$^{\#}$ \\ 
& [29] & \TEXTsymbol{>}2.6$\times 10^{17}$ & [33] & SU(4) & 0.198 & 13.00 & 
32.15 \\ \hline
ECEC & [30] & \TEXTsymbol{>}5.8$\times 10^{17}$ & [31] & PHFB & 0.238 & 11.24
& 28.42 \\ 
10$^{20}$yrs & \multicolumn{2}{|l|}{} & [32] & QRPA & 0.169 & 22.24$^{\#}$ & 
55.0$^{\#}$ \\ 
& \multicolumn{2}{|l|}{} & [33] & SU(4) & 0.193 & 17.00 & 42.04 \\ \hline
\end{tabular}

$^{*}$\textit{\ denotes half-life limit for 0}$\nu $\textit{\ + 2}$\nu $%
\textit{\ mode, }$^{**}$\textit{denotes half-life limit for 0}$\nu $\textit{%
\ + 2}$\nu $\textit{\ +0}$\nu $\textit{M mode, }$^{\#}$\textit{shows
half-life with WS potential, }$^{\dagger }$\textit{Number below the mode is
multiplication factor of Half-life for theoretical values.}

\textit{\smallskip }

\textit{\smallskip }From the overall agreement [16] between the calculated
and observed electromagnetic properties, it is clear that the PHFB wave
functions of $^{100}$Mo and $^{100}$Ru generated by fixing $\chi _{pn}$ to
reproduce the yrast spectra are quite reliable.

\smallskip The double beta decay of $^{100}$Mo $\rightarrow ^{100}$Ru for $%
0^{+}\rightarrow 0^{+}$ transition has been investigated by many
experimental groups [22,23] as well as theoreticians by employing different
theoretical frameworks [24-26]. In Table 2, we have compiled some of the
latest available experimental and the theoretical results along with our
calculated M$_{2\nu }$ and the corresponding half-life T$_{1/2}^{2\nu }$. We
have used a phase space factor G$_{2\nu }$ = 9.434 $\times $10$^{-18}$ yr$%
^{-1}$ given by Doi et al [2] and an energy denominator E$_{d}$ = 11.2 MeV
given by Haxton et al [1]. In column 4 of Table 2, we have presented the M$%
_{2\nu }$ extracted from the experimentally observed T$_{1/2}^{2\nu }$ using
the phase space factor given above. The phase space integral has been
evaluated for g$_{A}$=1.25 by Doi et al [2]. However in heavy nuclei it is
more justified to use the nuclear matter value of g$_{A}$ around 1.0. Hence,
the experimental M$_{2\nu }$ as well as the theoretical T$_{1/2}^{2\nu }$
are calculated for g$_{A}$=1.0 and 1.25. The present calculation and that of
Hirsch et al using SU3(SPH) [25] give nearly identical value. They are close
to the experimental result given by De Silva et al [22] for g$_{A}$=1.25
while for g$_{A}$=1.0, the above two M$_{2\nu }$ are in agreement with the
results of NEMO. The calculated values given by Stoica using SRPA(WS) [24]
are too low and those from Suhonen et al [26] are slightly on higher side.
Further the value M$_{2\nu }$ given by Hirsch et al using SU3(DEF) [25]
favors the results of NEMO [23] for g$_{A}$=1.25.

\smallskip Another example we take for the 0$^{+}\rightarrow $0$^{+}$%
positron $\beta \beta $ ($\beta ^{+}\beta ^{+}$, $\beta ^{+}$EC and ECEC)
decay of $^{106}$Cd$\rightarrow ^{106}$Pd. This transition has also been
investigated by many experimental groups and in different theoretical
frameworks. In Table 3, we have compiled some of the latest available
experimental [27-30] and the theoretical results [31-33] along with our
calculated M$_{2\nu }$ and corresponding half-lives T$_{1/2}^{2\nu }$ . We
have used phase space factors given by Doi and Kotani [3] and the average
energy from Haxton and Stephenson [1]. Our calculated values are nearly half
of the recently given QRPA results of Suhonen and Civitaresse [32] for all
the three modes. The theoretical values of PHFB and SU(4) [33] are in 
better agreement (factor of roughly two third) for the $\beta ^{+}$EC and
ECEC modes.

From the above discussions, it is clear that the validity of nuclear models
presently employed to calculate the M$_{2\nu }$ cannot be uniquely
established due to error bars in experimental results as well as uncertainty
in g$_{A}$. Further work is necessary both on the experimental and
theoretical front to judge the relative applicability, success and failure
of various models used so far for the study of double beta decay processes.

As an example to see quantitatively the effect of deformation on M$_{2\nu }$
vis-a-vis the variation of the strength of pn part of the QQ interaction,
the results are displayed in Fig.1 for $^{100}$Mo and $^{106}$Cd cases. It
is observed that the deformations of the HFB intrinsic states play an
important role in the calculations of M$_{2\nu }$ and hence on the half life.

To summarize, we have first tested the quality of HFB wave functions by
comparing the theoretically calculated results for a number of spectroscopic
properties of nuclei involved in double beta decay. To be more specific we
have computed the yrast spectra, reduced B(E2) transition probabilities,
quadrupole moments and g-factors. Some of the results have been presented
for two very widely studied cases of $\beta ^{-}\beta ^{-}$ decaying $^{100}$%
Mo and e$^{+}\beta \beta $ ($\beta ^{+}\beta ^{+}$, $\beta ^{+}$EC and ECEC)
decaying $^{106}$Cd nuclei. Reliability of the intrinsic wave functions for
calculation of 2$\nu $ $\beta \beta $ nuclear matrix elements, M$_{2\nu }$ ,
has been discussed. Further, we have shown that the np interactions viz a
viz the deformations of the intrinsic ground states of $^{100}$Mo $^{100}$%
Ru, $^{106}$Cd and $^{106}$Pd play important role in arriving at the
appropriate nuclear matrix elements. A reasonable agreement between the
calculated and observed spectroscopic properties as well as the 2$\nu $ $%
\beta \beta $ decay rate of most of the nuclei in medium mass region makes
us confident to employ the same PHFB wave functions for the study of 0$\nu $ 
$\beta \beta $ decay.

\end{document}